\documentclass[letter]{aa}

\usepackage{graphicx}
\usepackage{txfonts}
\usepackage{amsmath}
\usepackage{upgreek}

\newcommand{\um}{\,\mu\mathrm{m}}
\newcommand{\cmc}{\,\mathrm{cm^{-3}}}
\newcommand{\cmq}{\,\mathrm{cm^{-2}}}
\newcommand{\Msun}{\,\mathrm{M}_{\sun}}
\newcommand{\Lsun}{\,\mathrm{L}_{\sun}}
\newcommand{\Av}{A_V}
\newcommand{\dR}{\Delta R}
\newcommand{\Ncl}{N_{\mathrm{cloud}}}
\newcommand{\Npdr}{N_{\mathrm{PDR}}}
\newcommand{\Rs}{R_{\mathrm{S}}}

\begin{document}

\title{The universal thickness of mid-infrared bubbles}

\author{
    Michael M.~Schulreich\corrauth{schulreich@astro.physik.tu-berlin.de}
    \and
    A.~Beate C.~Patzer\email{patzer@astro.physik.tu-berlin.de}
    }

\institute{
    Institut f\"ur Physik und Astronomie, Technische Universit\"at Berlin,
    Hardenbergstra{\ss}e 36, 10623 Berlin, Germany
    }

\date{Received / Accepted}

\abstract
{Several hundred ring-like structures seen at $8\um$ in the \emph{Spitzer} GLIMPSE
surveys share a strikingly uniform morphology: the thickness $\dR$ of the emitting
shell is a nearly constant fraction of its outer radius $R$,
$\dR/R\sim0.3$, across a factor of 40 in size and across
all exciting stellar spectral types. No
explanation for this invariance has been given so far.}
{We investigated which physical quantity sets the observed shell thickness and whether the
near-constancy of $\dR/R$ is a genuine physical fingerprint or a measurement
artifact.}
{We derived purely analytic constraints on the location, temperature, and heating
of the emitting grains and compared the resulting predictions for the scaling of
$\dR/R$ with bubble radius against the combined GLIMPSE~I and~II bubble catalogs
(591 objects, 91 with kinematic distances).}
{Grains in radiative equilibrium at the observed radii only reach $20$--$35$\,K
and fall short of the required $8\um$ emissivity by more than $20$ orders of
magnitude;
the emission must be carried by stochastically heated polycyclic aromatic
hydrocarbons and very small grains, and it therefore originates in the
photon-dominated region, where the ring width is the far-ultraviolet
penetration depth. This gives $\dR/R=\Av^{\mathrm{PDR}}/\Av^{\mathrm{cloud}}$,
independent of radius. A swept-up shell instead predicts
$\dR/R\propto(R/\Rs)^{3/2}$, with $\Rs$ the initial Str\"omgren radius. We found a pronounced
angular-resolution systematic that inflates $\dR/R$ for the smallest bubbles;
once this is accounted for, $\dR/R=0.217$.
Its residual dependence on physical radius is
$0.03\pm0.05$, consistent with zero; a fixed ambient density is
excluded at $20\sigma$ and the swept-up shell at
$9$--$28\sigma$.
The implied ambient columns are
$\Ncl\approx8.7\times10^{21}\cmq$, with a scatter of only $0.112$\,dex.}
{The near-universal appearance of mid-infrared bubbles follows from the ratio of
two nearly invariant column densities: the dust-opacity-fixed far-ultraviolet
shielding column, and the Larson-like column of the parent molecular cloud. It
requires no reference to stellar wind properties.}


\maketitle
\nolinenumbers

\section{Introduction}
\label{sec:intro}

Visual inspection of the \emph{Spitzer} Galactic Legacy Infrared Mid-Plane Survey
Extraordinaire (GLIMPSE) revealed 322 partial and
closed $8\um$ rings in $10\degr<|l|<65\degr$ \citep{churchwell06} and a further
269 in the inner $20\degr$ \citep{churchwell07}. These so-called bubbles are now known
to enclose H\,\textsc{ii} regions in ${\gtrsim}85\%$ of cases
\citep{deharveng10,anderson11}, and about half of all Galactic H\,\textsc{ii}
regions show a bubble morphology \citep{anderson11}.

\citet{churchwell06} noted an empirical regularity that has attracted little
attention since: the ratio $\dR/R$ of the shell thickness to the outer radius
is tightly clustered between $0.2$ and $0.4$, with $\dR$ growing approximately
linearly with $R$.
The authors compared this with \citet{weaver77} models computed for a range of ambient
densities and two stellar spectral types, and they attributed the flatter observed trend to a
spread in evolutionary stage. They also cautioned that the models refer to the
swept-up gas shell, whereas the measurement refers to the dust
emission, which at $5.8$ and $8.0\um$ is dominated by aromatic bands and might
extend beyond the compressed layer \citep{churchwell07}. This Letter takes this
second remark as its starting point and makes it quantitative.
Sections~\ref{sec:constraints} and~\ref{sec:pdr} set out the two interpretations,
Sect.~\ref{sec:data} tests their differing radius scalings, and
Sects.~\ref{sec:disc} and~\ref{sec:concl} discuss our results and conclude.

\section{Origin of the $8\um$ emission}
\label{sec:constraints}

\subsection{The rings are not equilibrium dust continuum}
\label{sec:notequil}

A grain of radius $a$ at a distance $r$ from a star of luminosity $L$ attains a
steady-state temperature that is well approximated by
$T_{\mathrm{d}}\approx16.4\left(a/0.1\um\right)^{-1/15}U^{1/6}$\,K for silicate and
$T_{\mathrm{d}}\approx22.3\left(a/0.1\um\right)^{-1/40}U^{1/6}$\,K for graphite grains
\citep{draine11}, where $U$ is the radiation energy density in units of the local
interstellar value. For $L=(1$--$5)\times10^{4}\Lsun$, the range spanned by the
late-O and early-B stars that power these bubbles \citep{beaumont10}, these give $T_{\mathrm{d}}=20$--$35$\,K at
$r=2$\,pc;
\emph{Herschel} measurements of the photon-dominated regions (PDRs) of bubbles
give $T_{\mathrm{d}}\approx26$\,K with remarkably little variation between
objects \citep{anderson12}.

At these temperatures, $8\um$ lies far out on the Wien tail. Relative to a $1000$\,K
grain, $B_\nu(8\um)$ is suppressed by a factor ${\sim}10^{-29}$ at $26$\,K, ${\sim}10^{-12}$ at
$60$\,K, and still ${\sim}10^{-7}$ at $100$\,K. A detectable $8\um$ continuum
would require $T_{\mathrm{d}}\gtrsim200$\,K. Because $T_{\mathrm{d}}\propto U^{1/6}
\propto r^{-1/3}$, this means $r\lesssim0.011$\,pc even in the warmest
case, which is orders of magnitude below the observed ring radii.

The emission must therefore be non-equilibrium. The Infrared Array Camera (IRAC) band~4 contains the $7.7$ and
$8.6\um$ aromatic features, and \citet{churchwell06,churchwell07} already
attributed the rings to polycyclic aromatic hydrocarbons (PAHs). This assignment
was confirmed by AKARI and \emph{Herschel} spectral energy distributions that separate
the PAH, warm-dust, and cold-dust components in several hundred bubbles
\citep{hattori16,hanaoka19,hanaoka20}. In the single-photon
heating regime, the emitted spectral shape is set by the absorbed photon
energy distribution and the molecular mode structure, not by the field intensity:
the band emission efficiencies are independent of $U$ for $U\lesssim10^4$
\citep{draineli07}, a limit these objects are far from ($U\approx3$--$15$ at
$2$\,pc), so the band ratios are invariant while the surface
brightness scales linearly with the far-ultraviolet (FUV) field. This is the origin of the uniform
color of the bubbles.

Collisional heating is irrelevant here: for $n\sim10^{3}$--$10^{4}\cmc$ and
$T\sim10^{2}$--$10^{4}$\,K at $r\sim2$\,pc from an OB star, the ratio of
collisional to radiative heating of a $0.1\um$ grain is $10^{-6}$--$10^{-2}$.
Because $T_{\mathrm{d}}$ goes as the sixth root of the heating rate, even the
extreme case shifts it by less than $1\%$. Only in the hot shocked wind do electron
collisions become competitive \citep{everett10}, but that region emits at
$24\um$.

\subsection{What the swept-up shell would predict}
\label{sec:shell}

The natural alternative, implicitly adopted whenever $\dR/R$ is compared with
wind-bubble models \citep[e.g.,][]{churchwell06,churchwell07}, is that the ring
traces the layer of ambient gas collected by the expansion. Mass conservation for
a thin shell containing all matter interior to $R$ gives
\begin{equation}
\label{eq:thinshell}
   \frac{\dR}{R } = \frac{1}{3} \frac{n_0}{n_{\mathrm{sh}}} ,
\end{equation}
where $n_0$ is the ambient and $n_{\mathrm{sh}}$ the shell density.

It is important to recognize that $n_{\mathrm{sh}}/n_0$ is not a fixed shock jump
but declines as the region expands. A molecular layer at $T_{\mathrm{sh}}$,
pressure-confined by an ionized interior of density $n_{\mathrm{i}}$ and
temperature $T_{\mathrm{i}}$, has
$n_{\mathrm{sh}} = (\mu_{\mathrm{sh}}/\mu_{\mathrm{i}}) n_{\mathrm{i}}
T_{\mathrm{i}}/T_{\mathrm{sh}}$ when both densities count hydrogen nuclei, and for
D-type
expansion $n_{\mathrm{i}}\approx n_0 (\Rs/R)^{3/2}$, where
$\Rs = [3Q / (4\uppi \alpha_{\mathrm{B}}n_0^2)]^{1/3}$ is the initial Str\"omgren radius,
$Q$ is the stellar ionizing photon rate, and $\alpha_{\mathrm{B}}$ is the case-B
recombination coefficient. Equation~(\ref{eq:thinshell}) then becomes
\begin{equation}
\label{eq:shellevol}
   \frac{\dR}{R} = \frac{1}{3} \frac{\mu_{\mathrm{i}}}{\mu_{\mathrm{sh}}}
               \frac{T_{\mathrm{sh}}}{T_{\mathrm{i}}}
               \left(\frac{R}{\Rs}\right)^{\!3/2} .
\end{equation}
For ionized and molecular gas at $n(\mathrm{He})/n(\mathrm{H})=0.1$,
$\mu_{\mathrm{i}}=0.64$ and $\mu_{\mathrm{sh}}=2.33$; with $T_{\mathrm{sh}}=50$\,K
and $T_{\mathrm{i}}=10^4$\,K, $\dR/R$ runs from $0.0024$ at $R/\Rs=3$ through
$0.075$ at $R/\Rs=30$ to $0.1$ near $R/\Rs=36$, beyond which the
thin-shell approximation loses accuracy; $\dR/R$ then grows without bound as the
region approaches stalling. Equating the ionized-gas and ambient pressures
gives $R_{\mathrm{stall}}/\Rs=(c_{\mathrm{i}}/c_0)^{4/3}$, with
$c_{\mathrm{i}}$ and $c_0$ the isothermal sound speeds in the ionized and in the
undisturbed molecular gas. For an isothermal shock, the layer cools back to
ambient, $T_{\mathrm{sh}}=T_0$, so with $c_{\mathrm{i}}=11.4$ and
$c_0=0.42\,\mathrm{km\,s^{-1}}$, the shell stalls at $R/\Rs\approx81$, while
$\dR/R=0.3$ is reached at $R/\Rs\approx62$ (evaluated with the exact mass
conservation because Eq.~(\ref{eq:shellevol}) is no longer accurate at that
thickness): the observed value is attained at $76\%$ of the stalling radius, and this
fraction is independent of both the adopted temperature and the helium abundance.

The swept-up shell can therefore reproduce the observed value at one
particular stage. What it cannot reproduce is the observed
invariance: Eq.~(\ref{eq:shellevol}) demands that $\dR/R$ vary by
nearly three orders of magnitude over a bubble's lifetime. Since
$\Rs\propto(Q/n_0^2)^{1/3}$, the predicted scaling with radius is
$\dR/R\propto R^{3/2}n_0Q^{-1/2}$, that is,
$\mathrm{d}\log(\dR/R)/\mathrm{d}\log R=3/2$ for a fixed ambient density and
$1/2$ if $n_0\propto R^{-1}$. Both are tested in Sect.~\ref{sec:data}.

Projection does not modify this picture: the limb-brightened rim of an optically
thin spherical shell has a full width at half maximum of the order of $\dR$, so apparent
and true widths agree to within a factor of a few.

\section{The PDR interpretation}
\label{sec:pdr}

Section~\ref{sec:notequil} places the emitting material outside the ionization
front, in the FUV-illuminated skin of the ambient cloud, consistent with the
morphology established by \citet{watson08}: $24\um$ and $20$\,cm emission are
cospatial (the H\,\textsc{ii} region), and the $8\um$ emission switches on
immediately outside the ionization front.

The width of that skin is the depth at which the FUV field is extinguished.
When we write the shielding column as $\Npdr=1.87\times10^{21}\Av^{\mathrm{PDR}}\cmq$
with $\Av^{\mathrm{PDR}}\approx1$--$2$\,mag \citep{hollenbach99}, and when we assume that the
ambient medium is not strongly compressed on this scale (the collected layer is
filamentary rather than smooth, as \citealt{churchwell06} noted and as
instabilities in accelerating shells produce \citep[e.g.,][]{schulreich22}, so the
FUV leaks past it),
the ring thickness is
\begin{equation}
\label{eq:thickness}
   \dR = \frac{\Npdr}{n_0} .
\end{equation}
With the ambient column over the scale of the bubble identified as
$\Ncl\approx n_0R$, Eq.~(\ref{eq:thickness}) yields the central result of this Letter,
\begin{equation}
\label{eq:master}
   \frac{\dR}{R} = \frac{\Npdr}{\Ncl}
                 = \frac{\Av^{\mathrm{PDR}}}{\Av^{\mathrm{cloud}}} .
\end{equation}
The observed thickness ratio is thus a ratio of two column densities that are both close
to universal. The numerator is fixed by the dust opacity itself, a property of
interstellar dust rather than of any individual object; the denominator is
approximately constant when clouds share a common mean column, $n_0\propto R^{-1}$,
as Larson's third relation implies \citep{larson81} and as extinction measurements
confirm from cloud to cloud \citep{lombardi10}.

Together with Sect.~\ref{sec:shell}, this defines a set of mutually exclusive
predictions for $\beta_D\equiv\mathrm{d}\log(\dR/R)/\mathrm{d}\log R$, namely,
$\beta_D=0$ for PDR depth with a constant cloud column ($n_0\propto R^{-1}$;
model~A), $\beta_D=-1$ for PDR depth at fixed ambient density (model~B),
$\beta_D=1/2$ for a swept-up shell with $n_0\propto R^{-1}$ (model~C), and
$\beta_D=3/2$ for a swept-up shell at fixed ambient density (model~D).
Only models~A and~B share the PDR identification; models~C and~D test the
collected layer. Model~A is also the result given by a purely geometric self-similar shell
structure,
so the slope alone does not separate PDR depth from geometry; the two differ in
normalization, and we return to this in Sect.~\ref{sec:disc}.

\section{Comparison with the GLIMPSE catalogs}
\label{sec:data}

\begin{figure*}
\centering
\includegraphics[width=\hsize]{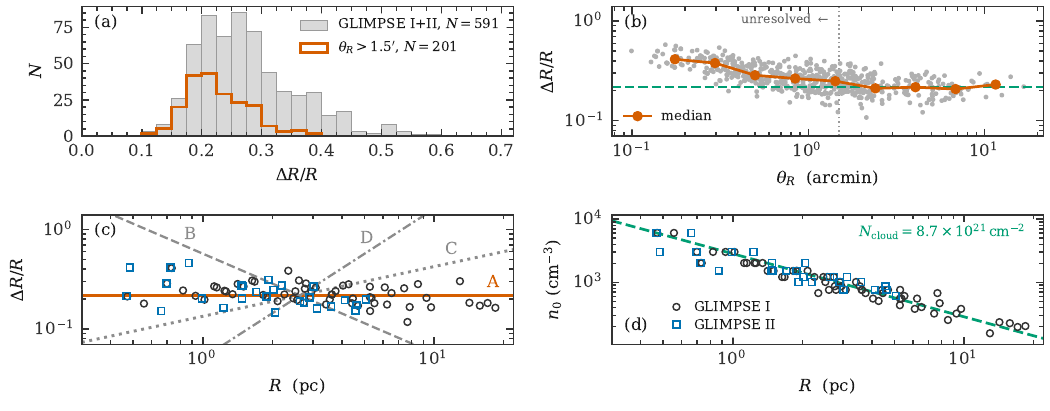}
\caption{(a) Distribution of $\dR/R$ for the combined GLIMPSE~I$+$II
sample (gray) and for the well-resolved subsample (orange). (b) Angular-resolution
systematic, with running medians; the dashed line marks the adopted intrinsic
value $0.217$. (c) $\dR/R$ against physical radius for the 91 bubbles with
kinematic distances, with models A--D of Sect.~\ref{sec:pdr} normalized at that
value; the symbols are the same as in panel~(d). (d) Ambient density implied by
Eq.~(\ref{eq:thickness}) for $\Av^{\mathrm{PDR}}=1$\,mag, compared with a
constant-column (Larson) locus.}
\label{fig:TR}
\end{figure*}

\subsection{Data}
We used the machine-readable catalogs of \citet[][322 bubbles;
CDS J/ApJ/649/759]{churchwell06} and \citet[][269 bubbles;
CDS J/ApJ/670/428]{churchwell07}, giving 591 objects with angular parameters.
Each bubble was described by concentric inner and outer ellipses with a common center
and eccentricity, whose geometric mean semiaxes we denote
$\bar R_{\mathrm{in}}$ and $\bar R_{\mathrm{out}}$. We took
$R\equiv\bar R_{\mathrm{out}}$, the boundary that carries the swept-up mass and
that is described by the expansion models, and
$\dR=\bar R_{\mathrm{out}}-\bar R_{\mathrm{in}}$, reproducing the
tabulated thickness to ${<}0\farcm02$. Kinematic near distances $D$, not part of the CDS
versions, were taken from table~4 of \citet{churchwell06} and table~3 of
\citet{churchwell07}. We discarded four entries whose tabulated $\dR$ rounds to
zero, which left 91 bubbles spanning $0.47\le R\le18.4$\,pc.

As a ratio of two angular quantities, $\dR/R$ is exactly independent of
distance. This is the key methodological advantage: distance errors move
objects horizontally in Fig.~\ref{fig:TR}c, but cannot generate a spurious
vertical trend.

\subsection{An angular-resolution systematic}
\label{sec:syst}
The combined sample gives a median $\dR/R=0.27$ with a scatter of
$0.124$\,dex
(Fig.~\ref{fig:TR}a).
However, $\dR/R$ correlates strongly with angular radius
$\theta_R$ (Fig.~\ref{fig:TR}b): the median rises from $0.22$ for
$\theta_R>3\arcmin$ to $0.44$ for $\theta_R<0\farcm2$, with
$\mathrm{d}\log(\dR/R)/\mathrm{d}\log\theta_R=-0.193\pm0.009$
($p=4\times10^{-71}$). Since the GLIMPSE point spread function is ${\sim}2\arcsec$
and the outer ellipse was placed by eye against a PDR that gradually fades outward
\citep{churchwell07}, this is a measurement systematic and not a physical
effect. It also accounts for the offset between the GLIMPSE~I and~II
distributions, the latter being dominated by smaller bubbles. For
$\theta_R>1\farcm5$, the trend flattens and $\dR/R=0.217$ with a scatter
of $0.097$\,dex. We adopted this as the intrinsic value. The 43 bubbles of
\citet{beaumont10}, selected on angular size, give $0.221$ with
$0.064$\,dex of scatter.

\subsection{Distinguishing the models}
Regression of $\log(\dR/R)$ on $\log R$ alone yields a slope
$-0.125\pm0.029$, which
is attributable to the systematic of Sect.~\ref{sec:syst}, however. Since
$\log R=\log\theta_R+\log D$, the coefficient of $\log D$ at fixed $\theta_R$ is
the derivative sought, and $\theta_R$ and $D$ are essentially uncorrelated in this
sample (Pearson $r=-0.16$, and variance inflation factor $1.03$), so the fit is well
conditioned. Table~\ref{tab:fit} gives $\beta_D=0.030\pm0.052$.

\begin{table}
\caption{Regression
$\log(\dR/R)=\alpha+\beta_\theta\log(\theta_R/1\arcmin)+\beta_D\log(D/\mathrm{kpc})$
for the 91 bubbles with
kinematic distances. The errors are $1\sigma$.}
\label{tab:fit}
\centering
\begin{tabular}{lr}
\hline\hline
Term & Coefficient \\
\hline
$\alpha$ (constant)                 & $-0.613\pm0.037$ \\
$\beta_\theta$ (systematic)         & $-0.165\pm0.030$ \\
$\beta_D$ (physical)                & $0.030\pm0.052$ \\
\hline
\multicolumn{2}{l}{\emph{Model predictions for $\beta_D$ (deviation)}}\\
A: PDR, $n_0\propto R^{-1}$               & $0$ ($0.6\sigma$) \\
B: PDR, $n_0= \mathrm{const}$             & $-1$ ($19.7\sigma$) \\
C: swept-up shell, $n_0\propto R^{-1}$    & $1/2$ ($9.0\sigma$) \\
D: swept-up shell, $n_0= \mathrm{const}$  & $3/2$ ($28.2\sigma$) \\
\hline
\end{tabular}
\end{table}

Model~A is recovered at $0.6\sigma$; the swept-up shell is
excluded at $9.0\sigma$ (model~C) and $28\sigma$ (model~D), and a fixed ambient
density is excluded at $20\sigma$ (model~B). The shell exponents $3/2$ and $1/2$ are
thin-shell limits: the exact relation steepens to $1.95$ and $0.65$ at the
observed thickness, so models~C and~D are tested against the weakest form of
their prediction. Restriction of the sample to better resolved bubbles confirms that the residual
slope is driven by the systematic: it relaxes from
$-0.125\pm0.029$ (all) to
$-0.091\pm0.039$ ($\theta_R>1\farcm5$) and
$-0.033\pm0.061$ ($\theta_R>2\farcm5$).

The shell model thus fails not on the value of $\dR/R$, which
Eq.~(\ref{eq:shellevol}) reaches close to stalling, but on its
constancy.
Reproducing the observed $0.097$\,dex scatter would require every bubble
to be caught within a factor ${\approx}1.7$ of the same $R/\Rs$, across
a factor of 40 in radius and the full range of exciting spectral
types. The spread in evolutionary
stage invoked by \citet{churchwell07} acts in the wrong direction: a range
$R/\Rs=20$--$60$ already implies $0.21$\,dex of scatter, which is more than twice
what is observed.

\subsection{Implied ambient columns}
Inversion of Eq.~(\ref{eq:thickness}) for each object with $\Av^{\mathrm{PDR}}=1$\,mag
gives a median $n_0=1010\cmc$ ($16$--$84\%$ range $550$--$2020\cmc$) and
$\Ncl=8.7\times10^{21}\cmq$, that is,
$\Av^{\mathrm{cloud}}\approx4.7$\,mag scaling linearly with
$\Av^{\mathrm{PDR}}$, with a scatter of only
$0.112$\,dex (Fig.~\ref{fig:TR}d). These match the values inferred for the molecular material around bubble
H\,\textsc{ii} regions \citep{deharveng10,anderson12} and are
comparable to the column at which a swept-up layer becomes
gravitationally unstable, $(4$--$8)\times10^{21}\cmq$, almost independently of the
driving mechanism \citep{whitworth94}.
Figure~\ref{fig:TR}d is not an independent test, since $n_0$ derives from $\dR$;
its value lies in making the prediction falsifiable.

\section{Discussion}
\label{sec:disc}

\subsection{A possible geometric rather than radiative invariance}
\label{sec:degeneracy}

A self-similar shell whose thickness is a fixed fraction of its radius
would reproduce $\beta_D=0$ equally well. Mass conservation ties this fraction
to the compression, however, and $\dR/R=0.217$ requires
$n_{\mathrm{sh}}/n_0\approx1.9$, which is produced by neither a radiative nor a
wind-driven shock (Appendix~\ref{app:selfsim}).

\subsection{Broken shells and formation scenarios}
\label{sec:broken}
Many bubbles do not show complete rings in projection
\citep[e.g.,][]{torii15,liu17,hattori16,hanaoka19}. A ruptured shell loses mass,
whereas a photodissociation front does not: its depth is set locally, normal to
the front. Using the covering fractions of \citet{hanaoka19}, we found $40$ closed
and $28$ broken bubbles with indistinguishable thickness ratios, $0.217$ against
$0.220$ (Appendix~\ref{app:cf}). Whether their asymmetries arise from
cloud--cloud collisions \citep{torii15,fukui21,kohno21}, champagne flows
\citep{ochsendorf14}, or anisotropic expansion \citep{luisi21}, $\dR/R$ is
untouched: it neither constrains such scenarios nor requires them.

\subsection{Predictions}
\label{sec:pred}

Two consequences can be tested with existing data. First, combining
Eq.~(\ref{eq:thickness}) with $\beta_D=0$ requires $n_0\propto R^{-1}$ at
$\Ncl\approx8.7\times10^{21}\cmq$ (Fig.~\ref{fig:TR}d); columns from the \emph{Herschel}
infrared Galactic Plane Survey (Hi-GAL; \citealt{molinari10}), the APEX
Telescope Large Area Survey of the Galaxy (ATLASGAL; \citealt{schuller09}),
or CO toward the same
bubbles must reproduce both slope and normalization; this is the sharpest available
falsification of the model. Second, since IRAC bands~1, 3, and~4 all
contain aromatic features and single-photon heating fixes the shape of that
spectrum but not its amplitude, the $[3.6]-[5.8]$ and $[5.8]-[8.0]$ colors should
be independent of exciting spectral type and of $R$.

\subsection{Caveats}
\label{sec:caveats}

The observed dispersion cannot yet constrain the model. From their four-examiner
experiment on 21 bubbles, \citet{churchwell06} quoted per-object dispersions
of about $35\%$ in thickness and $\dR/R=0.24\pm0.03$ across examiners. Our
resolved value, $0.217$ with $0.097$\,dex ($25\%$) of scatter, lies
below that uncertainty, so measurement error plausibly accounts for
the entire spread. A shielding column drawn uniformly from $1$--$2$\,mag would
contribute $0.087$\,dex, which is accommodated by the observed dispersion but is not
constrained by it.

Two further points should be kept in mind. First, the bubbles sit in
flattened clouds whose thickness is at most comparable to the ring diameter, so
they lack caps along the line of sight, at least for the 43 objects mapped in CO
by \citet{beaumont10}; as $\dR$ is measured at the limb, these do not enter, but
$\Ncl=n_0R$ is then within a factor of 2 of the column along the short axis in
the comparison with Larson-type columns. Less comfortably, the same authors derived a mean ring column of
$3\times10^{20}\cmq$ in hydrogen nuclei from CO, which is a factor of $30$ below
our $\Ncl$; but that column implies $\Av\approx0.16$\,mag, for which no
aromatic ring could form at all, so the CO values are likely underestimates. Second, our
interpretation concerns the $8\um$ rings only: the interior $24\um$ emission is
shaped by radiation pressure and grain drift \citep{ochsendorf15} and should not
be modeled with the same dust population.

\section{Conclusions}
\label{sec:concl}

\begin{enumerate}
\item Grains in radiative equilibrium at the observed bubble radii reach only
      $20$--$35$\,K and cannot produce the observed $8\um$ emission, falling short
      by more than $20$ orders of magnitude in emissivity. The rings are stochastically
      heated PAH and very-small-grain emission.
\item A swept-up, pressure-confined shell reaches the observed value of $\dR/R$
      only late in its life, and only by ceasing to be thin; it predicts
      $\dR/R\propto(R/\Rs)^{3/2}$ and is excluded by the absence of any such trend
      at $9$--$28\sigma$.
\item Identification of the ring with the FUV penetration depth of the ambient cloud
      gives $\dR/R=\Av^{\mathrm{PDR}}/\Av^{\mathrm{cloud}}$. This is the ratio of two nearly
      universal column densities.
\item The combined GLIMPSE~I$+$II catalogs contain a strong angular-resolution
      systematic in $\dR/R$. When this is removed, the intrinsic ratio is $\dR/R=0.217$ and is independent of the physical
      radius ($\beta_D=0.03\pm0.05$). This excludes a constant
      ambient density at $20\sigma$.
\item The implied ambient columns, $\Ncl\approx8.7\times10^{21}\cmq$
      with $0.112$\,dex scatter, match independent determinations
      without adjustment.
\end{enumerate}

\begin{acknowledgements}
We thank the anonymous referee, whose suggestions led to the test
presented in Appendix~\ref{app:cf}. 
This research has made use of the VizieR 
catalog access tool, CDS, Strasbourg, France (DOI: 10.26093/cds/vizier). 
\end{acknowledgements}

\bibliographystyle{aa}
\bibliography{bubbles}

@ARTICLE{anderson11,
       author = {{Anderson}, L.~D. and {Bania}, T.~M. and {Balser}, Dana S. and {Rood}, Robert T.},
        title = "{The Green Bank Telescope H II Region Discovery Survey. II. The Source Catalog}",
      journal = {\apjs},
         year = 2011,
        month = jun,
       volume = {194},
       number = {2},
          eid = {32},
        pages = {32},
          doi = {10.1088/0067-0049/194/2/32},
archivePrefix = {arXiv},
       eprint = {1103.5085},
 primaryClass = {astro-ph.GA},
       adsurl = {https://ui.adsabs.harvard.edu/abs/2011ApJS..194...32A}
}

@ARTICLE{anderson12,
       author = {{Anderson}, L.~D. and {Zavagno}, A. and {Deharveng}, L. and {Abergel}, A. and {Motte}, F. and {Andr{\'e}}, Ph. and {Bernard}, J.-P. and {Bontemps}, S. and {Hennemann}, M. and {Hill}, T. and {Rod{\'o}n}, J.~A. and {Roussel}, H. and {Russeil}, D.},
        title = "{The dust properties of bubble H II regions as seen by Herschel}",
      journal = {\aap},
         year = 2012,
        month = jun,
       volume = {542},
          eid = {A10},
        pages = {A10},
          doi = {10.1051/0004-6361/201117283},
archivePrefix = {arXiv},
       eprint = {1203.5721},
 primaryClass = {astro-ph.GA},
       adsurl = {https://ui.adsabs.harvard.edu/abs/2012A&A...542A..10A}
}

@ARTICLE{beaumont10,
       author = {{Beaumont}, Christopher N. and {Williams}, Jonathan P.},
        title = "{Molecular Rings Around Interstellar Bubbles and the Thickness of Star-Forming Clouds}",
      journal = {\apj},
         year = 2010,
        month = feb,
       volume = {709},
       number = {2},
        pages = {791-800},
          doi = {10.1088/0004-637X/709/2/791},
archivePrefix = {arXiv},
       eprint = {0912.1852},
 primaryClass = {astro-ph.GA},
       adsurl = {https://ui.adsabs.harvard.edu/abs/2010ApJ...709..791B}
}

@ARTICLE{churchwell06,
       author = {{Churchwell}, E. and {Povich}, M.~S. and {Allen}, D. and {Taylor}, M.~G. and {Meade}, M.~R. and {Babler}, B.~L. and {Indebetouw}, R. and {Watson}, C. and {Whitney}, B.~A. and {Wolfire}, M.~G. and {Bania}, T.~M. and {Benjamin}, R.~A. and {Clemens}, D.~P. and {Cohen}, M. and {Cyganowski}, C.~J. and {Jackson}, J.~M. and {Kobulnicky}, H.~A. and {Mathis}, J.~S. and {Mercer}, E.~P. and {Stolovy}, S.~R. and {Uzpen}, B. and {Watson}, D.~F. and {Wolff}, M.~J.},
        title = "{The Bubbling Galactic Disk}",
      journal = {\apj},
         year = 2006,
        month = oct,
       volume = {649},
       number = {2},
        pages = {759-778},
          doi = {10.1086/507015},
       adsurl = {https://ui.adsabs.harvard.edu/abs/2006ApJ...649..759C}
}

@ARTICLE{churchwell07,
       author = {{Churchwell}, E. and {Watson}, D.~F. and {Povich}, M.~S. and {Taylor}, M.~G. and {Babler}, B.~L. and {Meade}, M.~R. and {Benjamin}, R.~A. and {Indebetouw}, R. and {Whitney}, B.~A.},
        title = "{The Bubbling Galactic Disk. II. The Inner 20{\textdegree}}",
      journal = {\apj},
         year = 2007,
        month = nov,
       volume = {670},
       number = {1},
        pages = {428-441},
          doi = {10.1086/521646},
       adsurl = {https://ui.adsabs.harvard.edu/abs/2007ApJ...670..428C}
}

@ARTICLE{deharveng10,
       author = {{Deharveng}, L. and {Schuller}, F. and {Anderson}, L.~D. and {Zavagno}, A. and {Wyrowski}, F. and {Menten}, K.~M. and {Bronfman}, L. and {Testi}, L. and {Walmsley}, C.~M. and {Wienen}, M.},
        title = "{A gallery of bubbles. The nature of the bubbles observed by Spitzer and what ATLASGAL tells us about the surrounding neutral material}",
      journal = {\aap},
         year = 2010,
        month = nov,
       volume = {523},
          eid = {A6},
        pages = {A6},
          doi = {10.1051/0004-6361/201014422},
archivePrefix = {arXiv},
       eprint = {1008.0926},
 primaryClass = {astro-ph.GA},
       adsurl = {https://ui.adsabs.harvard.edu/abs/2010A&A...523A...6D}
}

@ARTICLE{deharveng09,
       author = {{Deharveng}, L. and {Zavagno}, A. and {Schuller}, F. and {Caplan}, J. and {Pomar{\`e}s}, M. and {De Breuck}, C.},
        title = "{Star formation around RCW 120, the perfect bubble}",
      journal = {\aap},
         year = 2009,
        month = mar,
       volume = {496},
       number = {1},
        pages = {177-190},
          doi = {10.1051/0004-6361/200811337},
archivePrefix = {arXiv},
       eprint = {0902.0903},
 primaryClass = {astro-ph.GA},
       adsurl = {https://ui.adsabs.harvard.edu/abs/2009A&A...496..177D}
}

@BOOK{draine11,
       author = {{Draine}, B.~T.},
        title = "{Physics of the Interstellar and Intergalactic Medium}",
    publisher = {Princeton University Press},
      address = {Princeton, New Jersey},
         year = 2011,
        pages = {290},
       adsurl = {https://ui.adsabs.harvard.edu/abs/2011piim.book.....D}
}

@ARTICLE{draineli07,
       author = {{Draine}, B.~T. and {Li}, Aigen},
        title = "{Infrared Emission from Interstellar Dust. IV. The Silicate-Graphite-PAH Model in the Post-Spitzer Era}",
      journal = {\apj},
         year = 2007,
        month = mar,
       volume = {657},
       number = {2},
        pages = {810-837},
          doi = {10.1086/511055},
archivePrefix = {arXiv},
       eprint = {astro-ph/0608003},
 primaryClass = {astro-ph},
       adsurl = {https://ui.adsabs.harvard.edu/abs/2007ApJ...657..810D}
}

@ARTICLE{everett10,
       author = {{Everett}, J.~E. and {Churchwell}, E.},
        title = "{Dusty Wind-blown Bubbles}",
      journal = {\apj},
         year = 2010,
        month = apr,
       volume = {713},
       number = {1},
        pages = {592-602},
          doi = {10.1088/0004-637X/713/1/592},
archivePrefix = {arXiv},
       eprint = {1003.0838},
 primaryClass = {astro-ph.SR},
       adsurl = {https://ui.adsabs.harvard.edu/abs/2010ApJ...713..592E}
}

@ARTICLE{fukui21,
       author = {{Fukui}, Yasuo and {Habe}, Asao and {Inoue}, Tsuyoshi and {Enokiya}, Rei and {Tachihara}, Kengo},
        title = "{Cloud-cloud collisions and triggered star formation}",
      journal = {\pasj},
         year = 2021,
        month = jan,
       volume = {73},
        pages = {S1-S34},
          doi = {10.1093/pasj/psaa103},
archivePrefix = {arXiv},
       eprint = {2009.05077},
 primaryClass = {astro-ph.GA},
       adsurl = {https://ui.adsabs.harvard.edu/abs/2021PASJ...73S...1F}
}

@ARTICLE{hanaoka19,
       author = {{Hanaoka}, Misaki and {Kaneda}, Hidehiro and {Suzuki}, Toyoaki and {Kokusho}, Takuma and {Oyabu}, Shinki and {Ishihara}, Daisuke and {Kohno}, Mikito and {Furuta}, Takuya and {Tsuchikawa}, Takuro and {Saito}, Futoshi},
        title = "{Infrared properties of Galactic infrared bubbles in the whole Galactic plane}",
      journal = {\pasj},
         year = 2019,
        month = feb,
       volume = {71},
       number = {1},
          eid = {6},
        pages = {6},
          doi = {10.1093/pasj/psy126},
       adsurl = {https://ui.adsabs.harvard.edu/abs/2019PASJ...71....6H}
}

@ARTICLE{hanaoka20,
       author = {{Hanaoka}, Misaki and {Kaneda}, Hidehiro and {Suzuki}, Toyoaki and {Kokusho}, Takuma and {Oyabu}, Shinki and {Ishihara}, Daisuke and {Kohno}, Mikito and {Furuta}, Takuya and {Tsuchikawa}, Takuro and {Saito}, Futoshi},
        title = "{Spatial distributions of dust components in Galactic infrared bubbles}",
      journal = {\pasj},
         year = 2020,
        month = feb,
       volume = {72},
       number = {1},
          eid = {5},
        pages = {5},
          doi = {10.1093/pasj/psz123},
       adsurl = {https://ui.adsabs.harvard.edu/abs/2020PASJ...72....5H}
}

@ARTICLE{hattori16,
       author = {{Hattori}, Yuki and {Kaneda}, Hidehiro and {Ishihara}, Daisuke and {Yamagishi}, Mitsuyoshi and {Oyabu}, Shinki and {Kondo}, Toru and {Suzuki}, Toyoaki and {Kokusho}, Takuma},
        title = "{Mid- and far-infrared properties of Spitzer Galactic bubbles revealed by the AKARI all-sky surveys}",
      journal = {\pasj},
         year = 2016,
        month = jun,
       volume = {68},
       number = {3},
          eid = {37},
        pages = {37},
          doi = {10.1093/pasj/psw028},
       adsurl = {https://ui.adsabs.harvard.edu/abs/2016PASJ...68...37H}
}

@ARTICLE{hollenbach99,
       author = {{Hollenbach}, D.~J. and {Tielens}, A.~G.~G.~M.},
        title = "{Photodissociation regions in the interstellar medium of galaxies}",
      journal = {Reviews of Modern Physics},
         year = 1999,
        month = jan,
       volume = {71},
       number = {1},
        pages = {173-230},
          doi = {10.1103/RevModPhys.71.173},
       adsurl = {https://ui.adsabs.harvard.edu/abs/1999RvMP...71..173H}
}

@ARTICLE{kohno21,
       author = {{Kohno}, Mikito and {Tachihara}, Kengo and {Fujita}, Shinji and {Hattori}, Yusuke and {Torii}, Kazufumi and {Nishimura}, Atsushi and {Hanaoka}, Misaki and {Yoshiike}, Satoshi and {Enokiya}, Rei and {Hasegawa}, Keisuke and {Ohama}, Akio and {Sano}, Hidetoshi and {Yamamoto}, Hiroaki and {Fukui}, Yasuo},
        title = "{CO observations toward the isolated mid-infrared bubble S44: External triggering of O-star formation by a cloud-cloud collision}",
      journal = {\pasj},
         year = 2021,
        month = jan,
       volume = {73},
        pages = {S338-S354},
          doi = {10.1093/pasj/psy109},
archivePrefix = {arXiv},
       eprint = {1809.00118},
 primaryClass = {astro-ph.GA},
       adsurl = {https://ui.adsabs.harvard.edu/abs/2021PASJ...73S.338K}
}

@ARTICLE{larson81,
       author = {{Larson}, R.~B.},
        title = "{Turbulence and star formation in molecular clouds.}",
      journal = {\mnras},
         year = 1981,
        month = mar,
       volume = {194},
        pages = {809-826},
          doi = {10.1093/mnras/194.4.809},
       adsurl = {https://ui.adsabs.harvard.edu/abs/1981MNRAS.194..809L}
}

@ARTICLE{liu17,
       author = {{Liu}, Hong-Li and {Figueira}, Miguel and {Zavagno}, Annie and {Hill}, Tracey and {Schneider}, Nicola and {Men'shchikov}, Alexander and {Russeil}, Delphine and {Motte}, Fr{\'e}d{\'e}rique and {Tig{\'e}}, J{\'e}r{\'e}my and {Deharveng}, Lise and {Anderson}, Loren D. and {Li}, Jin-Zeng and {Wu}, Yuefang and {Yuan}, Jing-Hua and {Huang}, Maohai},
        title = "{Herschel observations of the Galactic H II region RCW 79}",
      journal = {\aap},
         year = 2017,
        month = jun,
       volume = {602},
          eid = {A95},
        pages = {A95},
          doi = {10.1051/0004-6361/201629915},
archivePrefix = {arXiv},
       eprint = {1702.01924},
 primaryClass = {astro-ph.GA},
       adsurl = {https://ui.adsabs.harvard.edu/abs/2017A&A...602A..95L}
}

@ARTICLE{lombardi10,
       author = {{Lombardi}, M. and {Alves}, J. and {Lada}, C.~J.},
        title = "{Larson's third law and the universality of molecular cloud structure}",
      journal = {\aap},
         year = 2010,
        month = sep,
       volume = {519},
          eid = {L7},
        pages = {L7},
          doi = {10.1051/0004-6361/201015282},
archivePrefix = {arXiv},
       eprint = {1008.4251},
 primaryClass = {astro-ph.GA},
       adsurl = {https://ui.adsabs.harvard.edu/abs/2010A&A...519L...7L}
}

@ARTICLE{luisi21,
       author = {{Luisi}, Matteo and {Anderson}, Loren D. and {Schneider}, Nicola and {Simon}, Robert and {Kabanovic}, Slawa and {G{\"u}sten}, Rolf and {Zavagno}, Annie and {Broos}, Patrick S. and {Buchbender}, Christof and {Guevara}, Cristian and {Jacobs}, Karl and {Justen}, Matthias and {Klein}, Bernd and {Linville}, Dylan and {R{\"o}llig}, Markus and {Russeil}, Delphine and {Stutzki}, J{\"u}rgen and {Tiwari}, Maitraiyee and {Townsley}, Leisa K. and {Tielens}, Alexander G.~G.~M.},
        title = "{Stellar feedback and triggered star formation in the prototypical bubble RCW 120}",
      journal = {Science Advances},
         year = 2021,
        month = apr,
       volume = {7},
       number = {15},
        pages = {eabe9511},
          doi = {10.1126/sciadv.abe9511},
archivePrefix = {arXiv},
       eprint = {2104.04568},
 primaryClass = {astro-ph.GA},
       adsurl = {https://ui.adsabs.harvard.edu/abs/2021SciA....7.9511L}
}

@ARTICLE{molinari10,
       author = {{Molinari}, S. and {Swinyard}, B. and {Bally}, J. and {Barlow}, M. and {Bernard}, J.-P. and {Martin}, P. and {Moore}, T. and {Noriega-Crespo}, A. and {Plume}, R. and {Testi}, L. and {Zavagno}, A. and {Abergel}, A. and {Ali}, B. and {Andr{\'e}}, P. and {Baluteau}, J.-P. and {Benedettini}, M. and {Bern{\'e}}, O. and {Billot}, N.~P. and {Blommaert}, J. and {Bontemps}, S. and {Boulanger}, F. and {Brand}, J. and {Brunt}, C. and {Burton}, M. and {Campeggio}, L. and {Carey}, S. and {Caselli}, P. and {Cesaroni}, R. and {Cernicharo}, J. and {Chakrabarti}, S. and {Chrysostomou}, A. and {Codella}, C. and {Cohen}, M. and {Compiegne}, M. and {Davis}, C.~J. and {de Bernardis}, P. and {de Gasperis}, G. and {Di Francesco}, J. and {di Giorgio}, A.~M. and {Elia}, D. and {Faustini}, F. and {Fischera}, J.~F. and {Fukui}, Y. and {Fuller}, G.~A. and {Ganga}, K. and {Garcia-Lario}, P. and {Giard}, M. and {Giardino}, G. and {Glenn}, J. and {Goldsmith}, P. and {Griffin}, M. and {Hoare}, M. and {Huang}, M. and {Jiang}, B. and {Joblin}, C. and {Joncas}, G. and {Juvela}, M. and {Kirk}, J. and {Lagache}, G. and {Li}, J.~Z. and {Lim}, T.~L. and {Lord}, S.~D. and {Lucas}, P.~W. and {Maiolo}, B. and {Marengo}, M. and {Marshall}, D. and {Masi}, S. and {Massi}, F. and {Matsuura}, M. and {Meny}, C. and {Minier}, V. and {Miville-Desch{\^e}nes}, M.-A. and {Montier}, L. and {Motte}, F. and {M{\"u}ller}, T.~G. and {Natoli}, P. and {Neves}, J. and {Olmi}, L. and {Paladini}, R. and {Paradis}, D. and {Pestalozzi}, M. and {Pezzuto}, S. and {Piacentini}, F. and {Pomar{\`e}s}, M. and {Popescu}, C.~C. and {Reach}, W.~T. and {Richer}, J. and {Ristorcelli}, I. and {Roy}, A. and {Royer}, P. and {Russeil}, D. and {Saraceno}, P. and {Sauvage}, M. and {Schilke}, P. and {Schneider-Bontemps}, N. and {Schuller}, F. and {Schultz}, B. and {Shepherd}, D.~S. and {Sibthorpe}, B. and {Smith}, H.~A. and {Smith}, M.~D. and {Spinoglio}, L. and {Stamatellos}, D. and {Strafella}, F. and {Stringfellow}, G. and {Sturm}, E. and {Taylor}, R. and {Thompson}, M.~A. and {Tuffs}, R.~J. and {Umana}, G. and {Valenziano}, L. and {Vavrek}, R. and {Viti}, S. and {Waelkens}, C. and {Ward-Thompson}, D. and {White}, G. and {Wyrowski}, F. and {Yorke}, H.~W. and {Zhang}, Q.},
        title = "{Hi-GAL: The Herschel Infrared Galactic Plane Survey}",
      journal = {\pasp},
         year = 2010,
        month = mar,
       volume = {122},
       number = {889},
        pages = {314},
          doi = {10.1086/651314},
archivePrefix = {arXiv},
       eprint = {1001.2106},
 primaryClass = {astro-ph.GA},
       adsurl = {https://ui.adsabs.harvard.edu/abs/2010PASP..122..314M}
}

@ARTICLE{ochsendorf15,
       author = {{Ochsendorf}, B.~B. and {Tielens}, A.~G.~G.~M.},
        title = "{A bimodal dust grain distribution in the IC 434 H ii region}",
      journal = {\aap},
         year = 2015,
        month = apr,
       volume = {576},
          eid = {A2},
        pages = {A2},
          doi = {10.1051/0004-6361/201424799},
archivePrefix = {arXiv},
       eprint = {1501.02256},
 primaryClass = {astro-ph.GA},
       adsurl = {https://ui.adsabs.harvard.edu/abs/2015A&A...576A...2O}
}

@ARTICLE{ochsendorf14,
       author = {{Ochsendorf}, B.~B. and {Verdolini}, S. and {Cox}, N.~L.~J. and {Bern{\'e}}, O. and {Kaper}, L. and {Tielens}, A.~G.~G.~M.},
        title = "{Radiation-pressure-driven dust waves inside bursting interstellar bubbles}",
      journal = {\aap},
         year = 2014,
        month = jun,
       volume = {566},
          eid = {A75},
        pages = {A75},
          doi = {10.1051/0004-6361/201423545},
archivePrefix = {arXiv},
       eprint = {1404.2807},
 primaryClass = {astro-ph.SR},
       adsurl = {https://ui.adsabs.harvard.edu/abs/2014A&A...566A..75O}
}

@ARTICLE{schuller09,
       author = {{Schuller}, F. and {Menten}, K.~M. and {Contreras}, Y. and {Wyrowski}, F. and {Schilke}, P. and {Bronfman}, L. and {Henning}, T. and {Walmsley}, C.~M. and {Beuther}, H. and {Bontemps}, S. and {Cesaroni}, R. and {Deharveng}, L. and {Garay}, G. and {Herpin}, F. and {Lefloch}, B. and {Linz}, H. and {Mardones}, D. and {Minier}, V. and {Molinari}, S. and {Motte}, F. and {Nyman}, L.-{\r{A}}. and {Reveret}, V. and {Risacher}, C. and {Russeil}, D. and {Schneider}, N. and {Testi}, L. and {Troost}, T. and {Vasyunina}, T. and {Wienen}, M. and {Zavagno}, A. and {Kovacs}, A. and {Kreysa}, E. and {Siringo}, G. and {Wei{\ss}}, A.},
        title = "{ATLASGAL - The APEX telescope large area survey of the galaxy at 870 {\ensuremath{\mu}}m}",
      journal = {\aap},
         year = 2009,
        month = sep,
       volume = {504},
       number = {2},
        pages = {415-427},
          doi = {10.1051/0004-6361/200811568},
archivePrefix = {arXiv},
       eprint = {0903.1369},
 primaryClass = {astro-ph.GA},
       adsurl = {https://ui.adsabs.harvard.edu/abs/2009A&A...504..415S}
}

@ARTICLE{schulreich22,
       author = {{Schulreich}, Michael M. and {Breitschwerdt}, Dieter},
        title = "{The time-dependent Rayleigh-Taylor instability in interstellar shells and supershells, including the eROSITA bubbles}",
      journal = {\mnras},
         year = 2022,
        month = jan,
       volume = {509},
       number = {1},
        pages = {716-737},
          doi = {10.1093/mnras/stab2940},
archivePrefix = {arXiv},
       eprint = {2110.04166},
 primaryClass = {astro-ph.HE},
       adsurl = {https://ui.adsabs.harvard.edu/abs/2022MNRAS.509..716S}
}

@ARTICLE{torii15,
       author = {{Torii}, K. and {Hasegawa}, K. and {Hattori}, Y. and {Sano}, H. and {Ohama}, A. and {Yamamoto}, H. and {Tachihara}, K. and {Soga}, S. and {Shimizu}, S. and {Okuda}, T. and {Mizuno}, N. and {Onishi}, T. and {Mizuno}, A. and {Fukui}, Y.},
        title = "{Cloud-Cloud Collision as a Trigger of the High-mass Star Formation: a Molecular Line Study in RCW120}",
      journal = {\apj},
         year = 2015,
        month = jun,
       volume = {806},
       number = {1},
          eid = {7},
        pages = {7},
          doi = {10.1088/0004-637X/806/1/7},
archivePrefix = {arXiv},
       eprint = {1503.00070},
 primaryClass = {astro-ph.GA},
       adsurl = {https://ui.adsabs.harvard.edu/abs/2015ApJ...806....7T}
}

@ARTICLE{watson08,
       author = {{Watson}, C. and {Povich}, M.~S. and {Churchwell}, E.~B. and {Babler}, B.~L. and {Chunev}, G. and {Hoare}, M. and {Indebetouw}, R. and {Meade}, M.~R. and {Robitaille}, T.~P. and {Whitney}, B.~A.},
        title = "{Infrared Dust Bubbles: Probing the Detailed Structure and Young Massive Stellar Populations of Galactic H II Regions}",
      journal = {\apj},
         year = 2008,
        month = jul,
       volume = {681},
       number = {2},
        pages = {1341-1355},
          doi = {10.1086/588005},
archivePrefix = {arXiv},
       eprint = {0806.0609},
 primaryClass = {astro-ph},
       adsurl = {https://ui.adsabs.harvard.edu/abs/2008ApJ...681.1341W}
}

@ARTICLE{weaver77,
       author = {{Weaver}, R. and {McCray}, R. and {Castor}, J. and {Shapiro}, P. and {Moore}, R.},
        title = "{Interstellar bubbles. II. Structure and evolution.}",
      journal = {\apj},
         year = 1977,
        month = dec,
       volume = {218},
        pages = {377-395},
          doi = {10.1086/155692},
       adsurl = {https://ui.adsabs.harvard.edu/abs/1977ApJ...218..377W}
}

@ARTICLE{whitworth94,
       author = {{Whitworth}, A.~P. and {Bhattal}, A.~S. and {Chapman}, S.~J. and {Disney}, M.~J. and {Turner}, J.~A.},
        title = "{The preferential formation of high-mass stars in shocked interstellar gas layers}",
      journal = {\mnras},
         year = 1994,
        month = may,
       volume = {268},
       number = {1},
        pages = {291-298},
          doi = {10.1093/mnras/268.1.291},
       adsurl = {https://ui.adsabs.harvard.edu/abs/1994MNRAS.268..291W}
}

@ARTICLE{zavagno06,
       author = {{Zavagno}, A. and {Deharveng}, L. and {Comer{\'o}n}, F. and {Brand}, J. and {Massi}, F. and {Caplan}, J. and {Russeil}, D.},
        title = "{Triggered massive-star formation on the borders of Galactic H II regions. II. Evidence for the collect and collapse process around RCW 79}",
      journal = {\aap},
         year = 2006,
        month = jan,
       volume = {446},
       number = {1},
        pages = {171-184},
          doi = {10.1051/0004-6361:20053952},
archivePrefix = {arXiv},
       eprint = {astro-ph/0509289},
 primaryClass = {astro-ph},
       adsurl = {https://ui.adsabs.harvard.edu/abs/2006A&A...446..171Z}
}

\begin{appendix}
\nolinenumbers
\section{A self-similar shell as an alternative}
\label{app:selfsim}

Model~A is not the only way to obtain $\beta_D=0$: a self-similar shell whose thickness
is a fixed fraction of its radius would give the same null slope, and the
comparison of $\dR/R$ with wind-bubble models made by
\citet{churchwell06,churchwell07} implicitly appeals to such a picture. The slope
alone cannot separate the two hypotheses; the normalization can. Any shell that
has swept up the medium interior to $R$ is constrained by mass conservation. At $\dR/R=0.217$ the layer is not thin, so the exact
form replaces Eq.~(\ref{eq:thinshell}) and forces a contrast
\begin{equation}
\label{eq:chi}
   \frac{n_{\mathrm{sh}}}{n_0}=\left[1-\left(1-\frac{\dR}{R}\right)^{3}\right]^{-1}
   \approx1.9 ,
\end{equation}
that is, a layer barely denser than the gas it has swept up. This is hard to
reconcile with what is known about the collected material, detected in CO, in the
submillimeter continuum, and in \emph{Herschel} column density maps as dense
condensations of $10^2$--$10^4\Msun$ that are colder than the surrounding PDR and
fragment to form stars \citep{zavagno06,deharveng09}.
Moreover, the contrast is not a free constant. For a layer confined by the
ionized interior, $n_{\mathrm{sh}}/n_0=(\mu_{\mathrm{sh}}/\mu_{\mathrm{i}})
(T_{\mathrm{i}}/T_{\mathrm{sh}})(\Rs/R)^{3/2}$, which follows from pressure
balance alone and needs no thin-shell assumption. It falls as the region expands
-- from ${\approx}140$ at $R/\Rs=3$ to unity at stalling -- so $\dR/R$ must rise,
precisely the behavior excluded in Sect.~\ref{sec:data}.

The radiative interpretation makes the opposite kind of statement. It does not
adjust a shell property to match $\dR/R$, but identifies the thickness with a
length, $\Npdr/n_0$, whose numerator is a property of interstellar dust rather
than of any individual object. Given $\Av^{\mathrm{PDR}}\approx1$--$2$\,mag, the
measured $\dR/R$ then predicts
$\Av^{\mathrm{cloud}}\approx5$--$9$\,mag for the parent clouds -- an externally testable number rather than a fitted one, and one
that agrees with the columns measured toward these objects.

\section{The thickness ratio of broken bubbles}
\label{app:cf}

\citet{hanaoka19} fitted the $8\um$ shells of $247$ bubbles and quantified how
much of each ring is present through a covering fraction, $\mathrm{CF}$, the
angular fraction of the shell detected above their threshold; $\mathrm{CF}=1$
denotes a ring that is complete in projection. This is a different geometry from
the missing caps of Sect.~\ref{sec:caveats}, which concern the line of sight;
bubbles identified as such are preferentially seen with the flattened axis toward
us \citep{beaumont10}, so a ring can be complete in projection and still lack
caps. Their table~2 lists $\mathrm{CF}$ for the $68$ objects
with distance determinations, all of which are in the GLIMPSE catalogs. Their
radii are measured independently of \citet{churchwell06,churchwell07} and agree
with the tabulated mean radii to $0.7\%$ in the median, with $12\%$ scatter; we
use their covering fractions only, and keep $\dR$ and $R$ from the GLIMPSE
catalogs throughout.

The two pictures make opposite predictions. A swept-up shell that ruptures
vents its interior pressure and loses part of the collected mass, so its
thickness should depend on how much of the shell survives. A photodissociation
front has no such memory: $\Npdr$ is the column over which the FUV field is
extinguished along the local normal, and the gaps in a broken ring change which
directions are illuminated, not how deep the field penetrates where the shell is
present.

\begin{figure}
\centering
\includegraphics[width=\columnwidth]{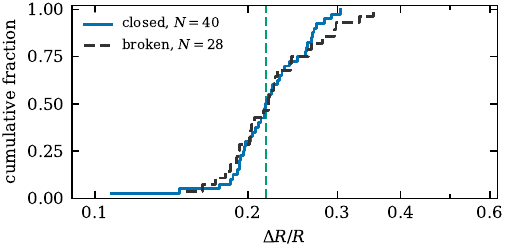}
\caption{Cumulative distributions of $\dR/R$ for closed and broken bubbles.
The dashed vertical line marks the adopted intrinsic value, $\dR/R=0.217$.}
\label{fig:cf}
\end{figure}

Table~\ref{tab:cf} and Fig.~\ref{fig:cf} show the outcome. Closed and broken
bubbles have the same thickness ratio to within one percent, and the two
distributions are statistically indistinguishable: a Kolmogorov--Smirnov test
gives $p=0.88$ and a Mann--Whitney test $p=0.94$. Regressing $\log(\dR/R)$ on
$\mathrm{CF}$ gives a slope of $-0.03\pm0.09$, consistent with zero. Restricting
to the well-resolved objects ($\theta_R>1\farcm5$, $N=51$) leaves the comparison
unchanged. The dispersions, $0.083$ and $0.089$\,dex, are if anything
smaller than in the resolved sample as a whole.

\begin{table}
\caption{Thickness ratio by covering fraction.}
\label{tab:cf}
\centering
\begin{tabular}{lccc}
\hline\hline
Subsample & $N$ & $\dR/R$ & scatter \\
\hline
$\mathrm{CF}=1.0$ (closed)    & $40$ & $0.217$ & $0.083$ \\
$\mathrm{CF}<1.0$ (broken)    & $28$ & $0.220$ & $0.089$ \\
\hline
$\mathrm{CF}<0.80$            & $15$ & $0.214$ & $0.106$ \\
$0.80\le\mathrm{CF}<0.95$     & $11$ & $0.225$ & $0.067$ \\
$\mathrm{CF}\ge0.95$          & $42$ & $0.217$ & $0.081$ \\
\hline
\end{tabular}
\tablefoot{Medians of $\dR/R$; scatter in dex. Covering fractions from
\citet{hanaoka19}, thicknesses and radii from
\citet{churchwell06,churchwell07}.}
\end{table}

This is a genuine test rather than a consistency check: the covering fractions
come from an independent analysis of independent imaging, and they span
$0.63\le\mathrm{CF}\le1$. That $\dR/R$ is blind to them supports the
identification of the ring with a radiatively determined depth, and disfavors
any mechanism in which the thickness records the accumulated mass.

The same insensitivity bears on the formation scenarios proposed for the
asymmetric objects. Cloud--cloud collisions
\citep[e.g.,][]{torii15,fukui21,kohno21} concentrate the gas into a compressed
layer on one side; champagne flows and punctured shells
\citep{ochsendorf14} drain ionized gas along the density gradient; and in
RCW\,120 the \ion{C}{ii} shell expands toward the observer only \citep{luisi21}.
All three
alter the global morphology, the covering fraction, and the distribution of
ionized gas, and \citet{hanaoka20} indeed find that broken bubbles differ
systematically from closed ones in the location of their heating sources and in
their hot-dust filling factors. None of them alters $\Ncl$, and $\Npdr$
only logarithmically: the depth to which the FUV penetrates grows as the
logarithm of its incident strength, so displacing the exciting star by $60\%$ of
the ring radius, the largest offset found by \citet{hanaoka20}, changes the
azimuthally averaged $\Av^{\mathrm{PDR}}$ by $0.3$\,mag for the standard
$A_{\mathrm{FUV}}/\Av=1.8$. The broken bubbles, which carry those offsets, show
neither a different $\dR/R$ nor a larger dispersion.
Eq.~(\ref{eq:master}) therefore predicts the same $\dR/R$ in all of them, as
observed: the thickness ratio gives no leverage on which scenario applies to a
given bubble, and none of them is needed to explain it.

\end{appendix}

\end{document}